\documentclass[aps,prl,preprint]{revtex4-2}
\usepackage{graphicx}
\usepackage{braket}
\begin{document}
\newcommand{\beq}{\begin{equation}}
\newcommand{\eeq}{\end{equation}}
\draft
\title  {Qubit States Based on Fractional Vortices in $0-\pi$ Josephson Junctions}
\author{ N. Stefanakis}
\affiliation{P.O. Box 1611, 72706, Reutlingen, Germany}

\date{\today}

\begin{abstract}
We investigate the static properties of $0-\pi$ Josephson junctions and their potential application as superconducting qubits. 
We show that fractional vortex and fractional antivortex configurations 
give rise to a bistable free-energy landscape with stable minima corresponding to topologically distinct fractional-vortex states. 
In the long-junction limit the two configurations correspond to stable equilibrium states, 
whereas in the short-junction limit the reduced barrier suggests enhanced quantum tunneling and hybridization between them.
These results suggest that $0-\pi$ Josephson junctions may provide a physically meaningful platform for qubit-like two-level dynamics, 
provided that a finite tunneling amplitude exists between the fractional-vortex and fractional-antivortex states.
\end{abstract}

\maketitle
\newpage

\section{Introduction}

Superconducting qubits have developed into one of the leading platforms for quantum information processing because they combine macroscopic quantum coherence, 
lithographic scalability, fast gate operations, and compatibility with microwave control and readout architectures. 
Their operation relies on the Josephson junction, which provides a nonlinear, 
nondissipative circuit element capable of transforming an otherwise harmonic LC oscillator into an anharmonic quantum system. 
This anharmonicity enables the isolation of two low-lying quantum states that can serve as a computational qubit basis.

Early superconducting-qubit implementations included charge qubits, flux qubits, and phase qubits, each emphasizing a different circuit degree of freedom. Charge qubits exploit the number of Cooper pairs on a superconducting island, flux qubits use persistent-current states in superconducting loops, and phase qubits rely on quantized levels in the tilted washboard potential of a current-biased Josephson junction. 
Although these early devices demonstrated coherent quantum behavior, their coherence was limited by sensitivity to charge noise, flux noise, dielectric defects, and other environmental perturbations.
For comprehensive reviews of superconducting quantum circuits see Refs.~\cite{devoret2013,blais2021,krantz2019}.

A major advance was the development of the transmon qubit \cite{koch2007}
which evolved from the Cooper-pair box by increasing the ratio between Josephson energy and charging energy. 
This design strongly suppresses charge-noise sensitivity while retaining sufficient anharmonicity for selective microwave control. 
As a result, transmon-type circuits, including 2D transmons, 3D transmons, Xmons, and related architectures, 
became central to many present-day superconducting quantum processors.

In parallel, flux-based qubits remained important because they provide a clear physical realization of a double-well energy landscape. 
In these systems, the two logical states are associated with clockwise and counterclockwise persistent currents. 
Near the degeneracy point, quantum tunneling between the two wells produces symmetric and antisymmetric superpositions, forming an effective two-level system. 
This double-well physics provides an important conceptual basis for understanding qubits based on alternative Josephson-junction configurations.

More recently, attention has turned toward protected superconducting qubits and hardware-level error suppression \cite{brooks2013}. 
Architectures such as fluxonium \cite{manucharyan2009}, capacitively shunted flux qubits, cat qubits \cite{mirrahimi2014}, and protected $0–\pi$ circuits aim to reduce decoherence by engineering the circuit Hamiltonian itself. 
In particular, the protected $0–\pi$ qubit proposed by Brooks, Kitaev, and Preskill uses circuit symmetry, superinductors, and nearly degenerate states to suppress relaxation and dephasing. 
These approaches are motivated by the need to reduce the overhead required for quantum error correction and to build more robust superconducting quantum hardware.

The present work follows this broader direction but focuses on a distinct physical realization: $0–\pi$ Josephson junctions with intrinsic phase discontinuities. 
Such junctions naturally support fractional vortex and fractional antivortex configurations. 
By solving the sine-Gordon equation under appropriate boundary conditions, we analyze the phase evolution and free-energy landscape of these systems and investigate whether they can provide a physical basis for qubit operation.

A central result of this work is that the energy-versus-flux dependence of the $0–\pi$ Josephson junction exhibits bistability between fractional vortex and 
fractional antivortex configurations. 
Although this behavior shares certain similarities with the double-well physics of superconducting flux qubits, 
the two states belong to distinct solution branches of the sine-Gordon equation and are selected by the preparation history of the junction.
The bistable free-energy landscape contains minima associated with
fractional-vortex and fractional-antivortex configurations.
The two branches become degenerate near zero applied flux. 
If a finite quantum tunneling amplitude exists between these topologically distinct configurations, 
symmetric and antisymmetric superpositions may be formed, providing a potential qubit basis.

This suggests that $0–\pi$ Josephson junctions can support qubit-like two-level dynamics based on intrinsic fractional-vortex physics.

This distinguishes the present proposal from conventional transmon and flux-qubit architectures. 
Instead of relying only on externally defined superconducting loops or large shunting capacitances, 
the qubit states arise from the internal phase structure of the Josephson junction itself. 
The intrinsic phase discontinuity of the $0–\pi$ junction therefore provides a natural mechanism for generating metastable states, double-well energy profiles, and potentially protected qubit configurations.

The aim of this paper is therefore to demonstrate that $0–\pi$ Josephson junctions can serve as physically meaningful building blocks for superconducting qubits. 
We analyze their static phase configurations, energy landscapes, and flux dependence, and we interpret the resulting fractional vortex and antivortex states as candidates for qubit-state encoding. 
The results support the view that $0–\pi$ Josephson junctions may offer a promising route toward protected Josephson-based qubits and motivate further studies of tunneling rates, coherence properties, microwave control, and environmental-noise effects.

\section{Josephson effect for $0-\pi$ junction} 

\begin{figure}[ht]
\centering
\includegraphics[width=0.65\linewidth]{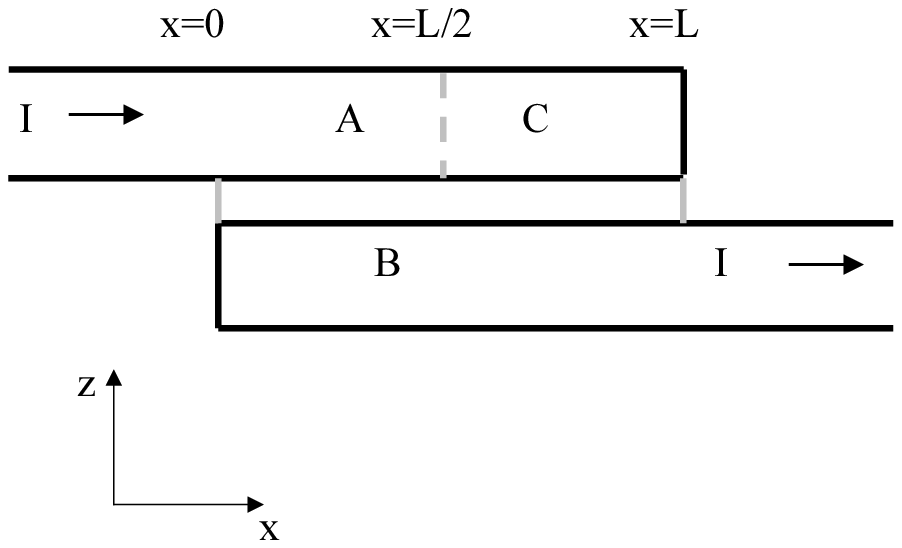}
\caption{
The Josephson junction of length $L$. 
}
\label{junction.fig}
\end{figure}

We describe the junction with width $w$ 
small compared to $\lambda_J$ in the $y$ direction, of length $L$ in the 
$x$ direction, in external magnetic field $H$ in the $y$ direction as shown in Fig. \ref{junction.fig}. 
The intrinsic phase difference  
$\theta(x)$ is $\phi_{c1}$ in $0<x<\frac{L}{2}$ and 
$\phi_{c2}$ in $\frac{L}{2}<x<L$, where $\phi_{c_1}=\phi_{c_2}=0$ for a $0-0$ junction 
and $\phi_{c_1}=0, \phi_{c_2}=\pi$ for a $0-\pi$ junction. 
The superconducting phase difference $\phi$ across the junction is 
then the solution of the Sine-Gordon equation
\beq 
  \frac{d^2 { \phi}(x)}{dx^2} = \frac{1}{\lambda_J^2}\sin[{
\phi(x)+\theta(x)}] ,~~~\label{eq01} \eeq with the inline boundary condition
\beq \frac{d { \phi}}{dx}\left|_{x=0,L}\right. =\pm \frac{{
I}}{2}+H   ~~~~\label{eq02} \eeq
The Josephson penetration depth is given by 
\[
\lambda_J=\sqrt{\frac{\hbar c^2}{8\pi e d \widetilde J_c}} 
\] where $d$ is the sum of 
the penetration depths in two superconductors plus the thickness of the 
insulator layer. We also assume that $\widetilde J_c$ is constant within 
each segment of the interface.

We compute the free energy
\beq 
  F = \frac{\hbar \widetilde J_c w}{2 e} \int_{0}^{L} \left[ 1-\cos 
\left[ \phi(x)+\theta(x) \right]
+\frac{\lambda_J^2}{2} \left( \frac{\partial \phi}{\partial x} \right)^2 \right] dx
 ,~~~\label{eq11} \eeq 

The free-energy landscape obtained from the sine-Gordon equation may serve as the basis
for a future quantum-mechanical treatment in which the corresponding Schrödinger equation
is solved to determine quantized energy levels and tunneling splittings.

Due to the $cos\phi$ dependence, the potential energy is anharmonic and the energy level splittings are 
unequal unlike the equal energy level splitting of the single harmonic oscillator or a superconducting LC circuit.
In a Josephson junction, the lowest energy levels can be addressed
using microwave excitation of energy $\hbar\omega_{01}$.

\section{Theory}
\subsection{Phase qubit} 
The phase qubit is a current-biased Josephson junction without any shunt capacitor or inductor. \cite{martinis2002}
The circuit can be considered by a current source, $I$, applied to a junction capacitance $C_J$ in parallel with a supercurrent $I_0\sin\phi$.
The potential energy is given by
\beq
U(\phi)=-E_J(\cos\phi+I\phi/I_0)
\eeq
where $E_J=\frac{\hbar I_0}{2e}=\frac{I_0 \Phi_0}{2\pi}$. This is an anharmonic potential. 
The energy levels in the potential well can be obtained by
solving the Schrödinger equation. The bias current $I$ can be adjusted so that only two energy levels are inside the well.
A microwave ac current can be used to drive oscillations between the states $\ket{0}$ und $\ket{1}$ forming a qubit.
The possibility of MQT switching to the resistive state is higher for the state $\ket{1}$. 
This makes possible to read the state of the qubit. 

\subsection{Flux qubit} 
The flux qubit is based on currents induced by an external magnetic field applied to a superconducting loop 
containing a Josephson junction. \cite{orlando1999,mooij1999}

The potential energy is given by
\beq
U(\phi)=-E_J\cos(2\pi\frac{\Phi}{\Phi_0})+\frac{(\Phi-\Phi_e)^2}{2L} 
\eeq
$U(\Phi)$ is a parabola modulated by a sinusoidal. Only the lowest energy in each well is important. 
When $\Phi_e>\Phi_0/2$ then the right potential well has lower energy 
favoring a counterclockwise circulating current 
that pushes the total flux to $\Phi_0$.
When $\Phi_e<\Phi_0/2$ then the left potential well has lower energy
favoring a clockwise circulating current 
that pushes the total flux to $0$.
When $\Phi_e=\Phi_0/2$ the energy levels are resonant. 
The degenerate energy levels at $\Phi_e=\Phi_0/2$ become coupled by quantum tunneling 
through the potential barrier between the two states. As a result of this coupling the 
degenerate energy levels are split. A superposition of clockwise and counter-clockwise 
supercurrent arises in the circuit.

\subsection{Charge Qubit}

The charge qubit, also known as the Cooper pair box, 
is a fundamental superconducting qubit architecture that utilizes the quantization of charge on 
a superconducting island connected to a reservoir via a Josephson junction. \cite{nakamura1999}
The system is controlled through a gate voltage applied via a gate capacitor, which allows precise tuning of the energy levels.

The Hamiltonian of the charge qubit is given by
\begin{equation}
H = 4E_C(n - n_g)^2 - E_J \cos\phi,
\end{equation}
where $E_C = \frac{e^2}{2C_\Sigma}$ is the charging energy, $E_J$ is the Josephson energy, $n$ is the number of excess Cooper pairs on the island, $n_g = \frac{C_g V_g}{2e}$ is the dimensionless gate charge, and $\phi$ is the superconducting phase difference across the junction.

The first term in the Hamiltonian represents the electrostatic energy due to the charge on the island, while the second term accounts for the tunneling of Cooper pairs through the Josephson junction. The interplay between these two energies allows the system to be operated as a quantum two-level system when biased near a degeneracy point ($n_g = 0.5$), where the energy levels are split due to tunneling and form an avoided crossing.

At this operating point, often referred to as the ``sweet spot,'' the qubit becomes less sensitive to charge noise, improving coherence properties. By applying microwave pulses at the qubit transition frequency, coherent oscillations (Rabi oscillations) can be induced between the ground and excited states, enabling qubit control.

Readout of the qubit state is typically achieved using a nearby charge sensor such as a single-electron transistor (SET), which detects the change in charge state of the qubit island. The Josephson energy $E_J$ can be made tunable by replacing the junction with a DC-SQUID, allowing additional control over the qubit dynamics.

\section{results}
In this section we solve numerically Eq. \ref{eq01} for different 
junction lengths and compute the Free Energy of the solutions.

\subsection{Flux qubit} 

We study first the case where the external current is $0$ and the 
magnetic field is modulated. 

We consider first the $0-0$ junction.
We plot in Fig. \ref{Energyl10_0-0.fig} the Free energy as a function of the flux $\Phi$ 
for the long length limit $L=10$.
We see that the Free energy corresponding to the different modes is a continuous function of the flux.

We consider next the case of a $0-\pi$ junction. 
The free energy as a function of magnetic flux for the long-junction limit $(L = 10)$ is shown in Fig. \ref{Energyl10_0-pi.fig}. 
In contrast to the $0-0$ case we see that here the different modes are overlapping.
The system can be prepared to the state with flux $0.5$ corresponding to the fractional vortex $fv$ 
or to the state with flux $-0.5$ corresponding to the fractional anti-vortex $fav$. 
Note that for zero flux the energy levels of the $fv$ and $fav$ branches become degenerate.
Whether coherent superpositions form depends on the existence of a finite quantum tunneling amplitude between the two branches.

One of the most important findings of the present work is the emergence of a bistable
free-energy landscape associated with fractional vortex and fractional antivortex states.
The calculated free-energy dependence on the applied magnetic flux demonstrates the existence of two energetically favorable configurations corresponding to fractional vortex ($fv$) and fractional antivortex ($fav$) states.

The $fv$ branch possesses a stable free-energy minimum near $\Phi=0.5$, whereas the $fav$ branch possesses a stable minimum near $\Phi=-0.5$. 
These minima demonstrate that the two configurations correspond to stable equilibrium states of the $0-\pi$ Josephson junction.

An important observation is that the $fv$ and $fav$ states belong to distinct solution branches of the sine-Gordon equation. 
The branch realized by the system depends on the preparation history and the initial phase configuration of the junction. 
Varying the external magnetic flux changes the free energy of each branch but does not continuously transform an $fv$ state into an $fav$ state.

Consequently, the observed bistability differs from that of a conventional superconducting flux qubit. \cite{orlando1999,mooij1999}
In a standard flux qubit, the external magnetic flux continuously controls the relative energies of two persistent-current states and tilts the double-well potential. 
In the present system, the $fv$ and $fav$ states should instead be interpreted as two topologically distinct fractional-vortex configurations.
The $fv$ and $fav$ states correspond to phase configurations with opposite fractional magnetic flux and opposite phase winding across the junction, 
making them topologically distinct solutions of the sine-Gordon equation.

At $\Phi=0$ the $fv$ and $fav$ branches become degenerate. 
If a finite quantum tunneling amplitude exists between these two branches, symmetric and antisymmetric superpositions of the form

\begin{equation}
|+\rangle=
\frac{1}{\sqrt{2}}
\left(
|fv\rangle+|fav\rangle
\right)
\end{equation}

\begin{equation}
|-\rangle=
\frac{1}{\sqrt{2}}
\left(
|fv\rangle-|fav\rangle
\right)
\end{equation}

may be formed. 
These states provide a natural qubit basis. 
The resulting qubit would therefore originate from tunneling between two topologically distinct fractional-vortex states 
rather than from conventional persistent-current states.

The energy-versus-flux dependence obtained here suggests that $0-\pi$ Josephson junctions may provide a platform for protected superconducting qubits based on fractional-vortex dynamics and engineered bistability.

In the small-junction limit $(L = 1)$, shown in Fig. \ref{Energyl1_0-0.fig}, 
the same qualitative picture remains valid. 
The $fv$ and $fav$ branches remain distinct and nearly degenerate around $\Phi=0$. 
However, the energy barrier separating the two branches is substantially reduced compared with the long-junction case. 
This suggests enhanced quantum tunneling and stronger hybridization between the two fractional-vortex configurations. 
The short-junction regime may therefore be particularly relevant for the realization
of coherent two-level dynamics if quantum tunneling between the $fv$ and $fav$ branches
is sufficiently strong.

\section{Relation to protected qubits and QEC}

The observed bistable free-energy structure is also relevant in the broader context of protected superconducting quantum circuits and quantum error correction. 
\cite{kitaev2003} Modern superconducting quantum-computing research increasingly investigates architectures that reduce
decoherence and error rates directly at the hardware level rather than relying exclusively on
active error correction.

In this context, the $0-\pi$ Josephson-junction system studied here shares conceptual similarities with protected superconducting qubits such as fluxonium circuits, cat qubits, and $0-\pi$
protected qubits. 
In all these architectures, engineered energy landscapes and metastable
quantum states are used to suppress unwanted transitions and improve coherence properties.

The existence of nearly degenerate $fv$ and $fav$ states in the present work indicates that
fractional-vortex configurations may serve as physically robust quantum states with reduced
sensitivity to external perturbations. The corresponding bistable free-energy landscape may
therefore provide a natural basis for future investigations of decoherence suppression, protected qubit operation, and hardware-efficient quantum error correction. Although the
present work focuses primarily on the physical realization of qubit-like states in $0-\pi$ Josephson junctions, the results suggest that further investigations involving coherence analysis,
tunneling rates, microwave control, and environmental coupling could establish deeper connections between fractional-vortex qubits and modern fault-tolerant superconducting quantum architectures.

\section{conclusions} 

The present work demonstrates that $0-\pi$ Josephson junctions naturally support stable fractional vortex ($fv$) and fractional antivortex ($fav$) configurations 
that generate a bistable free-energy landscape. 
In the long-junction limit, the two states correspond to well-defined free-energy minima near $(\Phi\approx \pm 0.5)$, 
while in the small-junction limit the barrier separating the two configurations is reduced, suggesting enhanced tunneling between them.

A key result of this work is that the $fv$ and $fav$ states belong 
to distinct solution branches of the sine-Gordon equation and are selected by the initial phase configuration of the junction. 
Consequently, the observed bistability differs from that of a conventional superconducting flux qubit, 
where the external magnetic flux continuously controls the relative energies of two persistent-current states. 
In the present system, varying the magnetic flux changes the energy of each branch but does not continuously transform an $fv$ state into an $fav$ state.

At zero applied magnetic flux the two branches become degenerate. 
If a finite quantum tunneling amplitude exists between the $fv$ and $fav$ configurations, 
symmetric and antisymmetric superpositions of these states may be formed, providing a natural qubit basis. 
The resulting qubit would therefore originate from tunneling between two topologically distinct fractional-vortex configurations.

These findings suggest that $0–\pi$ Josephson junctions may provide a promising platform for protected superconducting qubits 
based on fractional-vortex dynamics and motivate future investigations of tunneling rates, coherence properties, microwave control, 
decoherence mechanisms, and fault-tolerant quantum-computing architectures.

The reduced free-energy barrier observed in the short-junction limit
suggests enhanced tunneling and stronger hybridization between the $fv$
and $fav$ configurations.

A quantitative determination of the tunneling splitting and the
effective two-level Hamiltonian remains an important topic for
future work.

\begin{figure}[ht]
\centering
\includegraphics[width=0.65\linewidth]{Energyl10_0-0.eps}
\caption{
Energy of a $0-0$ junction of length $l=10$ versus the magnetic flux $\Phi$. 
The different modes correspond to different values of the magnetic flux
}
\label{Energyl10_0-0.fig}
\end{figure}

\begin{figure}[ht]
\centering
\includegraphics[width=0.65\linewidth]{Energyl10_0-pi.eps}
\caption{
Free-energy dependence of a $0–\pi$ Josephson junction with length $(l=10)$ as a function of magnetic flux $(\Phi)$. 
The fractional vortex $(fv)$ and fractional antivortex $(fav)$ branches possess stable minima near $(\Phi\approx +0.5)$ and $(\Phi\approx -0.5)$, respectively. 
The two branches correspond to distinct topological configurations selected by the initial phase distribution. 
At $(\Phi=0)$ the $fv$ and $fav$ states become degenerate, providing a potential operating point for quantum superpositions generated by tunneling between the two branches.}

\label{Energyl10_0-pi.fig}
\end{figure}

\begin{figure}[ht]
\centering
\includegraphics[width=0.65\linewidth]{Energyl1_0-0.eps}
\caption{
Energy of a $0-0$ junction of length $l=1$ versus the magnetic flux $\Phi$. 
}
\label{Energyl1_0-0.fig}
\end{figure}

\begin{figure}[ht]
\centering
\includegraphics[width=0.65\linewidth]{Energyl1_0-pi.eps}
\caption{
Free-energy dependence of a $0–\pi$ Josephson junction in the small-junction limit $(l=1)$. 
The fractional vortex $(fv)$ and fractional antivortex $(fav)$ branches remain bistable and nearly degenerate around $(\Phi=0)$. 
Compared with the long-junction case, the reduced barrier between the two branches suggests enhanced quantum tunneling 
and stronger hybridization between the corresponding fractional-vortex states.
}
\label{Energyl1_0-pi.fig}
\end{figure}

\end{document}